\documentclass[preprint,showkeys,a4paper]{revtex4-1}%%%%where rstrans is the template name

\usepackage{siunitx}
\usepackage{graphicx}

\begin{document}

%%%% Article title to be placed here
\title{Precision measurements on trapped antihydrogen in the ALPHA experiment}

\author{S. Eriksson}
\email{s.j.eriksson@swansea.ac.uk}
\affiliation{Department of Physics, College of Science, Swansea University, Singleton Park, Swansea SA2 8PP, UK}

%%%% Subject entries to be placed here %%%%
%\subject{Physics, Atomic Physics, High Energy Physics}

%%%% Keyword entries to be placed here %%%%

%%%% Insert corresponding author and its email address}
%\corres{S. Eriksson\\

%%%% Abstract text to be placed here %%%%%%%%%%%%
\begin{abstract}
Both the 1S-2S transition  and the ground state hyperfine spectrum have been observed in trapped antihydrogen. The former constitutes the first observation of resonant interaction of light with an anti-atom, and the latter is the first detailed measurement of a spectral feature in antihydrogen. Due to the narrow intrinsic linewidth of the 1S-2S transition and use of two-photon laser excitation, the transition energy can be precisely determined in both hydrogen and antihydrogen, allowing a direct comparison as a test of fundamental symmetry. The result is consistent with CPT invariance at a relative precision of around \num{2e-10}. This constitutes the most precise measurement of a property of antihydrogen. The hyperfine spectrum of antihydrogen is determined to a relative uncertainty of \num{4e-4}. The excited state and the hyperfine spectroscopy techniques currently both show sensitivity at the few \SI{100}{\kilo\hertz} level on the absolute scale. Here, the most recent work of the ALPHA collaboration on precision spectroscopy of antihydrogen is presented together with an outlook on improving the precision of measurements involving lasers and microwave radiation. Prospects of measuring the Lamb-shift  and determining the antiproton charge radius in trapped antihydrogen in the ALPHA-apparatus are presented. Future perspectives of precision measurements of trapped antihydrogen in the ALPHA apparatus when the ELENA facility becomes available to experiments at CERN are discussed. \end{abstract}
%%%%%%%%%%%%%%%%%%%%%%%%%%%

%%%%%%%%%% Insert the texts which can accomdate on firstpage in the tag "fmtext" %%%%%

%\begin{fmtext}

%\end{fmtext}

%%%%%%%%%%%%%%% End of first page %%%%%%%%%%%%%%%%%%%%%

\maketitle

\section{Introduction}
Antihydrogen, the antimatter equivalent of hydrogen, offers a unique way to test matter-antimatter symmetry. In particular, the CPT (charge, parity and time) theorem requires that hydrogen and antihydrogen have the same spectrum. Antihydrogen can reproducibly be synthesised and trapped in the laboratory for extended periods of time~\cite{Andresen2010,Andresen2011}, offering an opportunity to study the properties of antimatter in detail. New techniques to study antihydrogen have emerged; the ALPHA collaboration at CERN can now interrogate the ground state energy structure with resonant microwaves~\cite{Amole2012}, determine the gravitational mass to inertial mass ratio~\cite{Amole2013} and measure charge neutrality~\cite{Amole2014,Ahmadi2016}.  Very recently, the collaboration has observed the 1S-2S transition~\cite{Ahmadi2017}, and the ground state hyperfine spectrum~\cite{Ahmadi2017-2}. Thus, the collaboration has shown not just that the basic tools for precision spectroscopy of antihydrogen are available, but also provided a first, ground-breaking test of CPT-invariance with anithydrogen.

Measurements of the hydrogen spectrum together with its interpretation has a long and illustrious history which is intimately linked with the development of quantum mechanics. For example, the discovery that the 2S and 2P states in hydrogen do not have the same energy (now known as the Lamb shift)\cite{Lamb1947} is inextricably linked with the development of quantum electrodynamics (QED). Today, the 1S-2S transition in hydrogen is known with an uncertainty of only \SI{10}{\hertz} which corresponds to a relative uncertainty of \num{4e-15}~\cite{Parthey2011}. The ground state hyperfine interval is determined from experiments on the hydrogen maser\cite{Hellwig1970,Essen1971} with an uncertainty down to \SI{1}{\milli\hertz} corresponding to a relative uncertainty of \num{0.7e-12}. It is because such high accuracy can be achieved in hydrogen, and because on the whole the hydrogen atom is a well understood system, that  a comparison with antihydrogen is so compelling. However, in 2010  a measurement of the Lamb shift in muonic hydrogen (a bound state of a proton and a muon) yielded a  proton charge radius 5$\sigma$ smaller compared to the  CODATA value~\cite{Pohl2010}. This discrepancy is known as the "proton radius puzzle". A recent laser spectroscopic measurement of the Rydberg constant and proton size from atomic hydrogen is consistent with results from muonic hydrogen~\cite{Beyer2017}. However, a discrepancy remains between  laser spectroscopic measurements and measurements with other methods. 

The proton charge radius contributes approximately \SI{1.2}{\mega\hertz} to the Lamb shift of the 1S state (about \SI{8}{\giga\hertz})~\cite{Mohr2016}. Since both ground state and 2S state antihydrogen atoms are now available in experiments, and since the experimental frequency resolution in the 1S-2S experiment is already commensurate with the nuclear size correction, it is pertinent to ask whether the Lamb shift could be measured in antihydrogen.

In this article I describe how antihydrogen is synthesised, trapped and detected in the upgraded ALPHA-2 apparatus. Using the first two antihydrogen spectroscopy experiments as a guide, I present the prospects for a measurement of the Lamb shift using microwave spectroscopy in the manifold of the first excited state and excited state laser spectroscopy. I provide an outlook on the future of precision measurements with antiydrogen when extra low energy antiprotons become available at CERN.

\section{Experimental apparatus and trapping procedure}

\subsection{The ALPHA-2 apparatus}
\begin{figure}[!h]
\centering\includegraphics[width=\linewidth]{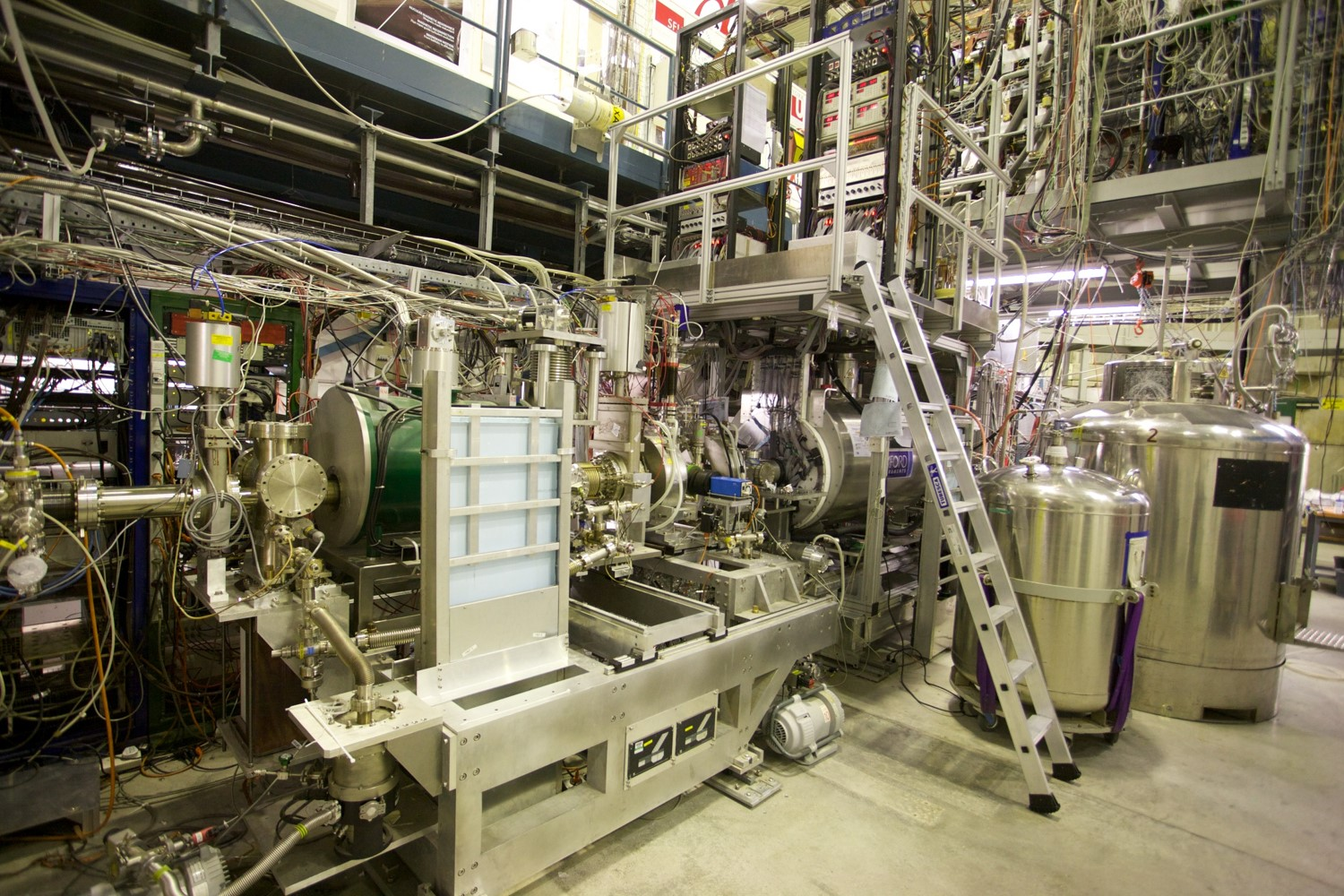}
\caption{The ALPHA-2 apparatus. The catching trap is on the antiproton beam line entering from the AD from the left. The main atom trap is located under the ladder-accessible platform which houses electronics for the annihilation vertex detector. The large dewar to the right of the atom trap contains liquid helium used to cool down the atom trap. The positron accumulator is on the beamline on the far right side of the atom trap, but is obscured by the dewars in this picture.}
\label{fig:figure1}
\end{figure}

The ALPHA experiment receives about 30,000,000 low energy (\SI{5.3}{\mega\electronvolt}) antiprotons from CERN's antiproton decelerator (AD) approximately every \SI{120}{\second} during daily 8-hour-long shifts throughout most of the antiproton run, which typically occurs during May-December. The antiprotons strike a beryllium degrader, and those which pass with the lowest energy are confined in a cryogenic Penning trap together with electrons which sympathetically cool down the antiprotons. In the upgraded ALPHA-2 apparatus (shown in figure~\ref{fig:figure1}) the antiproton capture operation takes place in a dedicated separate "catching trap" which is connected to the neighbouring main antihydrogen synthesis and trapping apparatus; the "atom trap". The superconducting magnet which provides the \SI{5}{\tesla} magnetic field for the catching trap can be seen in green in figure~\ref{fig:figure1}. The catching trap can be operated independently of the rest of the experiment. This feature allows optimisation of the antihydrogen production flux by  independently preparing the positrons in the positron accumulator while synchronising all necessary plasma manipulations to the AD deceleration cycle. 

\subsection{Antihydrogen synthesis and trapping}
\begin{figure}[!h]
\centering\includegraphics[width=\linewidth]{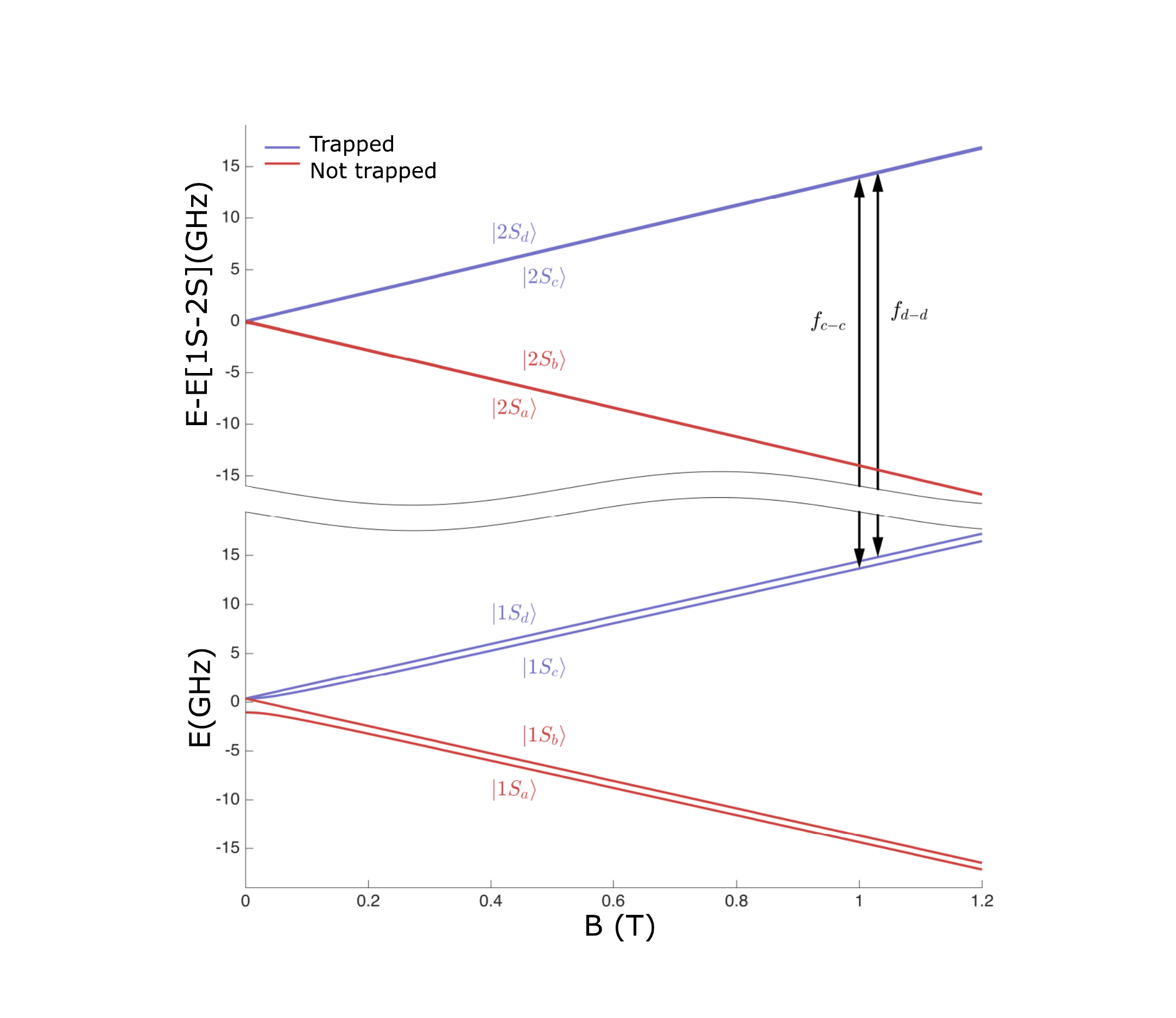}
\caption{Energy level diagram of the antihydrogen ground and first excited S-states in a magnetic field. The transitions indicated by arrows are discussed in section~\ref{section:laser}.}
\label{fig:figure2}
\end{figure}

Antihydrogen is synthesised in the atom trap by transferring the antiprotons from the catching trap and positrons from a Surko-style buffer gas accumulator~\cite{Surko2004}  into separate neighbouring potential wells. The charged plasmas are then merged to form antihydrogen in a process referred to as mixing while a minimum B-field magnetic trap is energised. Antihydrogen is formed in a three-body recombination process and  anti-atoms with lower kinetic energy than the trap depth of 0.5 K remain confined by interactions of their magnetic moment and the inhomogeneous field of the magnetic trap. Details of the synthesis process and trapping can be found e.g. in~\cite{Ahmadi2017-3}. Briefly, around 100,000 antiprotons and 2,000,000 positrons yield about 30,000 antihydrogen atoms per mixing attempt. Out of these, up to 20 are trapped in one production cycle which lasts \SI{4}{\minute}. The antihydrogen atoms form in an excited state, but decay rapidly to one of the four hyperfine states of the ground state manifold, labelled $a, b, c$ and $d$  with increasing energy, see figure~\ref{fig:figure2}. Of these four states $c$ and $d$ have the correct magnetic moment to become trapped.

\subsection{Detection by annihilation}

The presence of an antihydrogen atom in the trap is determined either by turning the magnetic trap off at the end of a trial or by manipulating the internal state of the atom in such away that it is no longer confined by the magnetic trap. The latter can be achieved resonantly by coupling microwaves to transitions between magnetic states within a manifold with one principal quantum number, $n$, or by coupling laser light to transitions that change $n$. In the case of two photon excitation (see section \ref{section:laser}) the laser is powerful enough to photoionise the antihydrogen atom from the 2S state by absorption of a third photon. When spin flips occur the entire anti-atom is ejected from the trap, and in the case of ionisation the antiproton is ejected. In both cases the antiproton annihilates on the Penning trap walls. The annihilation by-products (mostly high energy pions) create tracks in the three-layered silicon detector which surrounds the magnetic trap~\cite{Amole2013-vertex}. Tracks from one event (defined by the trigger mechanism of the detector) can be used to reconstruct a vertex which indicates the spatial location where the annihilation occurred~\cite{Andresen2012-annihilation}. A vertex coinciding with the Penning trap walls constitutes a signal from trapped antihydrogen. The detector is capable of detecting a single annihilation event with a spatial resolution along the axis of the trap of around \SI{7}{\milli\meter}. Combined, the spatial and temporal information yields information of the distribution of trapped antihydrogen atoms. The distributions can in turn be used to deduce the dynamics of anti-atoms whilst they were trapped, by comparing  to simulations of trapped anti-atoms~\cite{Amole2012-discriminating} .

\section{Two photon laser excitation}
\label{section:laser}
The 1S-2S transition in hydrogen and antihydrogen can be probed with two photons of half the excitation energy each, corresponding to a wavelenght of \SI{243}{\nano\meter}. The Doppler shift due to the residual motion of the antihydrogen atoms in the trap is cancelled to first order when the two photons are counter-propagating. For hydrogen atoms in a gas cell the transition rate scales with the laser intensity squared and inversely with the linewidth taking into account broadening effects. The motion of antihydrogen atoms in the trapping potential complicates this scaling somewhat, necessitating simulations for a refined prediction of the excitation rate~\cite{Rasmussen2017}. The excitation can be followed by photo-ionisation with another \SI{243}{\nano\meter} photon or a spin-flip. Both processes lead to ejection of the anti-atom from the magnetic trap, and a signal on the annihilation detector.

The laser system consists of a custom made Toptica FHG-pro laser which produces about \SI{100}{\milli\watt} of \SI{243}{\nano\meter} light by frequency doubling in two separate steps (FHG in the name of the laser refers to fourth harmonic generation), starting from a fundamental wavelength of \SI{972}{\nano\meter}. An approximately \SI{0.9}{\meter} long enhancement cavity situated in the cryogenic, ultra-high vacuum space provides build-up of circulating laser power to $p_l$=\SI{1}{W} with a beam-waist of 
$w$=200~$\mu$m
%\SI{200}{\micro\meter}
.
The extended cavity diode laser which produces the fundamental is frequency stabilised by coupling the light into a high-finesse optical resonator. The length of the resonator has an ultra low expansion rate, leading to a very low drift rate of the resonance peaks. The frequency drift is continuously monitored by detecting the beat note between the laser and a frequency comb which is referenced to atomic time via a GPS disciplined quartz oscillator. The frequency drift is continuously compensated for by shifting the laser frequency using an acousto-optic modulator which is also used to set the laser frequency during the antihydrogen spectroscopy efforts. The laser irradiation is kept constant for \SI{300}{\second} at each of the two possible two-photon transitions, $c\to c$ and $d \to d$, as shown in figure~\ref{fig:figure2}. In 2016, three types of trials were conducted: runs with the laser tuned to the expected hydrogenic resonance taking into account the residual Zeeman effect and the diamagnetic shift in the \SI{1}{\tesla} background field, runs with the laser detuned by \SI{-200}{\kilo\hertz} (at \SI{243}{\nano\meter}) from resonance, and control runs with the experimental condition identical to the two laser runs, but with the laser off. It was shown in~\cite{Ahmadi2017} that 58\% of the trapped antihydrogen atoms are removed when the laser is resonant, which is in good agreement with simulations~\cite{Rasmussen2017}. Comparing the frequency detuning to the optical frequency of the transition one reaches a relative precision of \num{2e-10}.

The obvious improvement of this first demonstration of resonant interaction with laser light and antihydrogen is to conduct a further measurement with the laser detuned by a positive detuning, thus showing providing a symmetric bound to the 1S-2S linewidth. 
A natural extension is to add data-points at finer frequency detuning intervals to allow detailed studies of the lineshape.  
These measurements can be performed with the current apparatus with no changes, and given the current antihydrogen trapping rates, such measurements are feasible during one  antiproton run or part thereof. With current antihydrogen synthesis and trapping methods, and without further cooling of the anti-atoms, the most significant linewidth broadening effect is due to the short time that the anti-atom spends traversing the laser beam. The approximate size of the broadening contribution of an anti-atom  in single-pass thought the laser beam is \SI{160}{\kilo\hertz}. In simulations, which take into account the distribution of kinetic energies, the estimated width becomes \SI{80}{\kilo\hertz} (at the transition frequency)~\cite{Rasmussen2017}. The simulations further show that a linewidth of approximately \SI{40}{\kilo\hertz} (at the transition frequency) could be achieved by increasing the laser beam waist to 400~$\mu$m.
%$\SI{400}{\micro\meter}
Such a reduction necessitates a change in the enhancement cavity geometry which would be challenging due to the large radius of curvature that the mirrors would be required to have in order to create a stable cavity, but nevertheless this is feasible with current technology. 
To minimise the residual Zeeman shift the $d\to d$ transition can be targeted by ejecting the $c$ state atoms from the trap with resonant microwaves. Thus, it is anticipated that with currently known techniques, a resonance lineshape with a width of order a few  \SI{10}{\kilo\hertz} can be experimentally determined. 
Assessing how accurately the 1S-2S transition line centre can be determined from these results is premature. 
Techniques currently under development such as laser cooling of antihydrogen, combined with adiabatic cooling will eventually reduce the line broadening to the few \SI{}{\kilo\hertz} level at which the AC Stark shift and broadening due to photoionisation is relevant. Such broadening could be avoided by resonant ejection by spin-flips in the 2S state or transitions between 2S and 2P states using microwaves. It is worth noting that with a lineshape measurement still outstanding, the simulations cannot yet be compared with data, and estimates given here should only be taken as guideline. 

\section{Ground state hyperfine spectroscopy}
\label{section:microwave}
Transitions between hyperfine states can occur without changing the principal quantum number, $n$. Within the 1S manifold the allowed transitions correspond to positron spin flips from $c \to b$ and $d \to a$ (with increasing frequency), with a separation corresponding to the hyperfine splitting of about \SI{1420}{\mega\hertz}. At the trap-centre field of \SI{1}{\tesla} these transitions can be driven resonantly with microwaves at about \SI{29}{\giga\hertz}. Since the $a$ and $b$ states are rapidly ejected from the trap, a clear signal is obtained when the microwaves are resonant. Due to the relatively flat magnetic field profile in the centre of the octupole-trap, the interaction is most likely to occur at a field magnitude corresponding to the bottom of the potential well. This leads to a rapid onset of the spin-flip signal when the frequency of the microwaves is swept from below resonance. The resonance shape has a long tail due to the inhomogeneous trapping field, but the sharp feature due to the rapid onset allows a precise determination of the hyperfine splitting. The trap provides a large number of ground state antihydrogen atoms compared to current alternative methods working towards in-beam measurements of the hyperfine splitting~\cite{Enomoto2010}. 

Microwaves at \SI{160}{\milli\watt} and \SI{320}{\milli\watt} at the lower and upper transitions, respectively, are injected via a waveguide reaching the central part of the trapping apparatus. The trapping structures act as waveguides for the microwaves and the higher power at the upper transition is chosen to partly compensate for the poor wave-guiding at that frequency. Measurements of cyclotron resonance heating of electron plasmas~\cite{Amole2014-plasma} indicate that the microwaves are about seven times weaker at the upper transition frequency, but increasing the power further leads to adverse thermal effects in the apparatus. The microwave frequency is detuned below the lower transition resonance and then stepped up in \SI{0.3}{\mega\hertz} intervals covering a total span of \SI{4.5}{\mega\hertz} in such a way that the onset of the spin-flip signal is observed. The frequency is then increased by the hyperfine interval and the process repeated for the upper transition. It was shown in~\cite{Ahmadi2017-2} that the resolution of the spectrum allows a determination of the hyperfine interval as \SI{1420.4\pm0.5}{\mega\hertz}.

While more detailed studies are necessary to fully understand the observed microwave spectrum, one may, based on the good agreement between the measured spectrum and simulations, speculate that the somewhat less rapid onset of the upper transition (compared to the lower) is simply due to the lower microwave power in the upper frequency. Using new trapping methods such as antihydrogen accumulation (see section \ref{section:accumulation}) one can envisage using more anti-atoms in the experiment and thus improving the signal. More anti-atoms should  allow a reduction of the current \SI{128}{s} microwave irradiation window, or alternatively the frequency step size. A reduction in  time of the irradiation window compensates for drifts in the magnetic field during the course of the microwave experiment. It can thus be anticipated that a significant improvement on the measurement of the hyperfine interval can be achieved, but a better estimate of the achievable precision requires optimising the experimental conditions. Such optimisation needs guidance from further simulations.

\section{Antihydrogen accumulation}
\label{section:accumulation}
Very recently it was shown that antihydrogen from up to five consecutive production cycles can be retained in the trap, yielding 54 trapped antihydrogen atoms in a single release~\cite{Ahmadi2017-3}. Thus, it is now possible to accumulate antihydrogen in the trap. Improved thermal control of the magnet system allows accumulation for longer times, making it feasible to consider experiments with thousands of anti-atoms at a time. While the current synthesis and accumulation technique does not increase the antihydrogen flux by much, this technique has far-reaching consequence on measurements due to the gain in signal-to-noise. For example, one can envisage exciting only a fraction of the ground state anti-atoms to the 2S state and performing excited state spectroscopy with similar signal-to-noise ratio as in the current laser and microwave experiments. We now turn to consider this scenario. 

\section{Prospects for a measurement of the Lamb-shift}
\label{section:Lamb}
There is clear evidence of laser excitation to the 2S state in the 1S-2S two-photon experiment conducted in the ALPHA-2 apparatus. Despite the reduction of lifetime from the natural lifetime of \SI{122}{\milli\second} to \SI{50}{\milli\second} due to the  motion of the atoms in a magnetic field, the 2S state is long-lived enough to consider excited state spectroscopy both in the optical and microwave domain. Below, transitions from 2S to 2P$_{1/2}$ and 2S to 4S are considered. 

\subsection{Microwave-driven 2S-2P transitions}

Atoms can be ejected from the trap by driving electric dipole allowed transitions from the trappable 2S state to one of the nearby 2P states (not shown in figure~\ref{fig:figure2}) which decay rapidly to an untrapped ground state. The waveguide structure in ALPHA-2 limits the microwave frequency to the range \SIrange[range-phrase = --]{22}{30}{\giga\hertz} which at the \SI{1}{\tesla} trapping field only leave the transitions to 2P$_{3/2}(m_j=3/2)$ and 2P$_{1/2}(m_j=-1/2)$. The former state never decays to an untrapped state and is therefore not useful when looking for an annihilation signal during the microwave irradiation. Coincidentally, at \SI{1}{\tesla} the two transition frequencies overlap at approximately \SI{24}{\giga\hertz}.  In ref.~\cite{Rasmussen2017}, it was found that decreasing the magnetic field to \SI{0.94}{\tesla} (which is within the apparatus capability) separates the transitions, and with \SI{1}{\milli\watt\per\square\centi\meter} at a frequency of \SI{22.5}{\giga\hertz} yields a transition rate of about \SI{2e4}{\hertz} for the transition to 2P$_{1/2}(m_j=-1/2)$, which decays to an untrapped state. Since it is not known when, during the 1S-2S irradiation, the anti-atom is excited, the microwaves would need to be pulsed at regular intervals of a few \SI{}{\milli\second} throughout, while looking for annihilation events. The magnetic field poses a systematic effect which can be eliminated by measurement. At the chosen background field, the resonance frequency shifts by about \SI{14}{\giga\hertz\per\tesla}. Since the magnetic field can be determined with $\Delta B/B=$\num{3.4e-4} from electron cyclotron resonance heating~\cite{Amole2014-plasma}, one can determine the resonance condition to \SI{4}{\mega\hertz}. Provided that the signal has a rapid onset akin to what is observed in the ground state microwave experiments, it can be anticipated that the transition frequency can be determined with a precision limited by the \SI{100}{\mega\hertz} linewdith the 2P state. Based on these simple considerations, a measurement of the Lamb shift in the 2S state could be pursued in the ALPHA-2 apparatus without modification as a first crude, unique test of QED in antihydrogen, keeping in mind that no such test exists to date. However, the two-photon excitation to 2S is followed by very efficient photo-ionisation, and obtaining a signal at all becomes a balance between the rate of ionisation and the microwave transition rate. One can reduce ionisation by lowering the intensity of the 1S-2S laser excitation, and to mitigate the reduced rate of excitation to the 2S state one can initially accumulate antihydrogen for longer. Detailed simulations of the microwave signal will be necessary to optimise the experimental protocol and parameters. 

Achieving a signal from a 2S-2P transition would indicate internal state control within the $n=2$ manifold,  which in turn would help guide future considerations on whether separated oscillatory field measurements~\cite{Lundeen1986,Hagley1994}, which achieve narrower frequency resolution at the expense of signal, are possible in trapped antihydrogen, or by using trapped antihydrogen as a source in a magnetically guided anti-atomic beam. Since the interference signal decreases exponentially with half the decay constant of the 2P state, and since very large microwave power is needed to create an appreciable Rabi-frequency, this is out of reach with current technology.

\subsection{Two-photon 2S-4S transitions}
\label{subsection:2S-4S}
The ground state Lamb shift can be determined by knowing the 1S-2S transition frequency together with a measurement of one other transition (see e.g.~\cite{Karshenboim2001}), preferably to a high $n$S  with a negligible Lamb shift. To assess the feasibility of a resonant 1S-2S laser excitation followed by spectroscopy of the 2S-4S line in trapped antihydrogen with two 972 nm photons, the following method is applied: it assumed that the laser beam of the 972 nm laser has the same geometry as the 243 nm laser (see section \ref{section:laser}). Such an arrangement is in principle feasible in ALPHA-2 with modest additions, since the 243 nm enhancement cavity mirrors are transparent to 972 nm light. The excitation probability is then estimated by simple scaling of the perturbative treatment in~\cite{Rasmussen2017} using the two-photon matrix element $\xi$ (here, in \SI{}{\hertz\per(\watt\per\meter\squared)}) and linewidth $\gamma$ of the 2S-4S transition in hydrogen from~\cite{Weitz1995,Haas2006}. According to equation (34) in~\cite{Rasmussen2017} the excitation probability during a single pass through the laser beam on resonance is $P= 4\pi^2 \xi^2 p_l^2 /(w^3 \delta f v)$ , where $\delta f$ is the total line broadening. With an antihydrogen velocity of $v$=\SI{90}{\meter\per\second}, and a negligible laser linewidth  $\delta f_{1S-2S}=v/(2\pi w)$=\SI{70}{\kilo\hertz}, for the 1S-2S transition, $P_{1S-2S}$=0.001. From table  II in~\cite{Haas2006}, the  transition matrix element is $\xi_{2S-4S}$=\SI{7.79e-5}{\hertz\per(\watt\per\meter\squared)}. Assuming that the only dominant broadening effect of the 2S-4S line is the natural linewidth, $\delta f_{2S-4S}$=\SI{706}{\kilo\hertz}, one finds that $P_{2S-4S}=1$ with $p_l$=\SI{47}{\watt}. Thus, an antihydrogen atom in the 2S state falling into the laser beam will be excited to the 4S state, and is very likely photoionised by a  third photon from the same laser. Since the arrival time of a 2S state atom into the laser beam is not known, the 2S-4S signal must be searched for in coincidence with an excitation pulse of a length commensurate with the lifetime of the 2S state. The necessary laser power could be generated with modest enhancement of a few Watts, readily available at \SI{972}{\nano\meter}. Since the enhancement is only modest, and since light at infra-red wavelengths is much more forgiving to mirrors and optics than \SI{243}{\nano\meter}, an enhancement cavity could be constructed external to the apparatus. While encouraging, this simple treatment neglects that the 1S-2S laser efficiently ionises the atom. Even though the ionisation coefficient for 2S-4S is an order of magnitude higher than for 1S-2S, there will be a background on the annihilation detector due to the 1S-2S laser. The unwanted ionisation rate can be lowered by decreasing the intensity at the expense of the number of excited state atoms, and as in the microwave case mitigated by accumulating more antihydrogen atoms. Detailed simulations of the optically induced signal including the motion of the antihydrogen atoms will be needed to provide guidance for the experiment. Based upon the \SI{5}{\kilo\hertz} estimate of the 1S-2S transition, the high 2S-4S intensity leads to an AC-Stark shift of order a few \SI{}{\mega\hertz}. To what extend this can be eliminated by extrapolation will need to await experimental verification of the signal, and knowledge about the rate at which data can be taken (see also section \ref{section:ELENA}). Thus, a test of bound state QED in antihydrogen which gives access to precision at the level of the contribution of the antiproton charge radius could be conducted in the optical domain, provided that a technical solution can be found for generating sufficient laser power at \SI{972}{\nano\meter}. It should be noted that at a magnetic field of \SI{1}{\tesla} the Zeeman energy dominates over the fine structure energy by at least one order of magnitude in the $n=4$ manifold. The effect on the 4S level by nearby $L\neq0$ states needs to be carefully assessed.

%An intriguing alternative would be to probe the single photon 2S-4P transition at \SI{486}{\nano\meter}~\cite{Berkeland1995}. However, in a gas, at a temperature corresponding an antihydrogen velocity $v=$\SI{90}{\meter\per\second} the Doppler broadening is about \SI{260}{\mega\hertz}.				

\section{Precision spectroscopy of antihydrogen in the ELENA era}
\label{section:ELENA}
In total the ALPHA-2 experiment has trapped around 50,000 antihydrogen atoms. The hyperfine spectrum measurement, which was conducted in 2016 over a period of three days with 22 trials, can be considered the state-of-the-art in data rate. The 2016 synthesis process results in about 11-14 trapped atoms per trapping cycle. CERN's Extra Low ENergy Antiproton ring, ELENA~\cite{ELENA2014}, is designed provide antiprotons to experiments with a kinetic energy of \SI{100}{\kilo\electronvolt}, and operation can in principle provide all (currently six, including ALPHA) experimental areas with antiprotons throughout the day, throughout the antiproton run. In addition to the 8-hour long antiproton shifts, ALPHA currently operates two 8-hour shifts with work on leptons. While a few hours of the lepton-shifts per day are needed for preparing the leptons for the antiproton shifts, one could envisage dedicating most of the 24-hour day to measurements with antihydrogen. One can thus immediately expect a nearly threefold increase in the current data rate with the arrival of ELENA. In principle, 2016-style laser spectroscopy campaign (11 trials with about 14 anti-atoms per trial) could be conducted yielding at least one new data-point in the spectrum per day. The effect is clearly beneficial, especially when the precision of measurements improves and studying systematic effects becomes harder. For example, the AC-Stark shift discussed in sections~\ref{section:laser} and~\ref{section:Lamb} could be investigated as a function of laser power. This simple consideration does not take into account that a significant effort was needed to stabilise the experiment before the spectroscopic measurements could begin. Having access to 24-hour antiprotons will speed up the annual start-up, and thus make more of the beam-period available for data collection. Thus, it becomes feasible to consider half-year long measurement campaigns which could produce sidereal constraints on Lorentz-violation. The lower energy of the antiprotons will ease trapping in the catching trap which can then be used as a reservoir, and thereby completely decouple the AD and antihydrogen production cycles.

\section{Conclusions}

Antihydrogen spectroscopy in the ALPHA-experiment was presented together with an outlook on the improvement of precision based upon known techniques. Using simple arguments it was shown that it is possible to measure the Lamb shift in trapped antihydrogen both in the 2S state directly and by excited state laser spectroscopy, with the latter showing promise to resolve the contribution of the antiproton charge radius. However, more detailed studies of the interaction between radiation and the antihydrogen atom in the complex trapping fields will be necessary to guide the experimental protocol. An outlook was given on the future of antihydrogen spectroscopy with low energy antiprotons  form ELENA become available. During the past year results from the ALPHA experiment have shown that precision spectroscopy is possible. Now it is imperative to seize upon this capability and explore every accessible aspect of the antihydrogen spectrum.

\vskip6pt

\enlargethispage{20pt}

%\ethics{Not applicable.}

%\dataccess{Not applicable.}

%\aucontribute{Not applicable.}

%\competing{The author(s) declare that they have no competing interests.}

%\funding{SE's work was funded by the EPSRC and the Leverhulme Trust.}

\acknowledgements{SE's work was funded by the EPSRC and the Leverhulme Trust. The author thanks Michael Charlton and Jeffrey Hangst for comments on the manuscript.}

%\disclaimer{Not applicable.}

%%%%%%%%%% Insert bibliography here %%%%%%%%%%%%%%


%\bibliography{references}
%\bibliographystyle{apsrev4-1}

\end{document}